\documentclass[a4paper,12pt]{article}
\usepackage{amsmath}
\usepackage{amsthm}
\usepackage{amssymb}

\title{\bfseries New Expressions for Discrete Painlev\'e Equations}
\author{By\\
Mikio Murata\\
(The University of Tokyo, Japan)}
\date{}

\theoremstyle{plain}
\newtheorem{thm}{Theorem}

\theoremstyle{definition}

\theoremstyle{remark}
\newtheorem{rem}{\slshape Remark}[section]

\numberwithin{equation}{section}

\begin{document}
\maketitle

\section{Introduction}
Discrete Painlev\'e equations are studied from various points of view as
integrable systems \cite{GORS}, \cite{RGH}.
They are discrete equations 
which are reduced to the Painlev\'e differential equations 
in a suitable limiting process,
and moreover, 
which pass the singularity confinement test. 
Passing this test can be thought of as a difference version of the Painlev\'e 
property.
The Painlev\'e differential equations were derived 
as second order ordinary differential equations 
whose solutions have no movable singularities 
other than poles. 
This property is called the Painlev\'e property.
The singularity confinement test has been proposed 
by Grammaticos \textit{et al.}\ as a criterion for the integrability 
of discrete dynamical systems \cite{GRP}. 
It demands that  
singularities depending on particular initial values 
should disappear after a finite number of iteration steps, 
in which case 
the information about the initial values ought to be recovered.

H. Sakai constructed all difference Painlev\'e 
equations from the point of view as algebraic geometry \cite{S}.
Surfaces obtained by successive blow-ups of 
$\mathbb{P}^1\times\mathbb{P}^1$ 
have been studied by means of connections between Weyl groups and
groups of Cremona isometries of the Picard group of surfaces.
The Picard group of a rational surface $X$ is the group of isomorphism
classes of invertible sheaves on $X$ 
and is isomorphic to the group of linear equivalence classes of
divisors on $X$.
A Cremona isometry is an isomorphism of the Picard group that 
i) preserves the intersection number of any pair of divisors,
ii) preserves the canonical divisor $\mathcal{K}_X$ 
and iii) leaves the set of effective classes of divisors invariant.
In the case where eight points in general positions are blown up in 
$\mathbb{P}^1\times\mathbb{P}^1$, 
the group of Cremona isometries of $X$
is isomorphic to an extension of the Weyl group of type $E_8^{(1)}$.
Birational mappings on $\mathbb{P}^1\times\mathbb{P}^1$ are obtained
by subsequent blow downs.
Discrete Painlev\'e equations are recovered as the birational mappings
corresponding to translations of affine Weyl groups.

Discrete Painlev\'e equations were classified on the basis of the 
types of rational surfaces 
connected to extended affine Weyl groups
and some new equations were discovered in the process. 
See Table~\ref{fig:class}.

\begin{table}[htbp]
\caption{Classification of generalized Halphen surfaces with 
$\dim\left|-\mathcal{K}_X \right|=0$}
\begin{center}
\begin{tabular}{c|l}
\hline
type & \multicolumn{1}{c}{surface (symmetry)}\\
\hline
Elliptic type & $A_0^{(1)}(E_8^{(1)})$\\
Multiplicative type & 
$A_0^{(1)*}(E_8^{(1)})\ A_1^{(1)}(E_7^{(1)})\ 
A_2^{(1)}(E_6^{(1)})\ A_3^{(1)}(D_5^{(1)})$\\ 
& $A_4^{(1)}(A_4^{(1)})\ A_5^{(1)}((A_2+A_1)^{(1)})\ 
A_6^{(1)}((A_1+\underset{|\alpha|^2=14}{A_1})^{(1)})$\\
& $A_7^{(1)}(\underset{|\alpha|^2=8}{A_1^{(1)}})\ 
A_7^{(1) \prime}(A_1^{(1)})\ A_8^{(1)}(A_0^{(1)})$\\
Additive type & 
$A_0^{(1)**}(E_8^{(1)})\ A_1^{(1)*}(E_7^{(1)})\ A_2^{(1)*}(E_6^{(1)})$\\
& $D_4^{(1)}(D_4^{(1)})\ D_5^{(1)}(A_3^{(1)})\ D_6^{(1)}((2A_1)^{(1)})$\\
& $D_7^{(1)}(\underset{|\alpha|^2=4}{A_1^{(1)}})\ D_8^{(1)}(A_0^{(1)})$\\
& $E_6^{(1)}(A_2^{(1)})\ E_7^{(1)}(A_1^{(1)})\ E_8^{(1)}(A_0^{(1)})$\\
\hline
\end{tabular}
\end{center}
\label{fig:class}
\end{table}

These equations are organized in a
degeneration pattern obtained through coalescence.
The $A_0^{(1)}$-surface discrete Painlev\'e equation 
($dP(A_0^{(1)})$) 
is the most generic one among these, 
because the equation has Weyl group symmetry of type $E_8^{(1)}$
and any of the other discrete Painlev\'e equations 
can be obtained from this equation by limiting procedure.
The form of this equation is however very complicated 
as its coefficients are described
in terms of elliptic functions \cite{MSY}, \cite{ORG}.

Kajiwara \textit{et al.}\ presented a $\tau$ function formalism 
for the elliptic discrete Painlev\'e equation \cite{KMNOY}.
They gave an explicit form of $dP(A_0^{(1)})$ based on this formalism.

In this paper 
a new representation of $dP(A_0^{(1)})$
based on total transforms is presented.
Kajiwara \textit{et al.}'s form is described in terms of coordinates 
in $\mathbb{P}^2$, 
while our form is described by means 
of $\mathbb{P}^1\times\mathbb{P}^1$.
Both descriptions are related by a bitational transform \cite{MSY}.
However, as in our construction all eight points 
(that are blown up in $\mathbb{P}^1\times\mathbb{P}^1$)
appear in a symmetric way,
the symmetries of $dP(A_0^{(1)})$ are immediately apparent.

In Section \ref{a0}, 
we present the representation of $dP(A_0^{(1)})$.
In Section \ref{a0*}--\ref{a3}, 
we present expressions for other discrete Painlev\'e equations, 
obtained in a similar way.
This is made possible by the fact that the limit process is a projective 
transformation of dependent variables.

Consequently, it can be seen that knowledge of the blow-up of
eight points in $\mathbb{P}^1\times\mathbb{P}^1$
and of a trivial solution on a curve passing through these points
allows one to construct a discrete Painlev\'e equation 
in explicit form.

\section{$A_0^{(1)}$-surface}\label{a0}
In this section, 
we give a new representation of the $A_0^{(1)}$-surface discrete 
Painlev\'e equation ($dP(A_0^{(1)})$).
First we present some useful notation.

We construct the $A_0^{(1)}$-surface by blowing up 
$\mathbb{P}^1\times\mathbb{P}^1$ at eight points 
$p_i\, (i=1,\dots,8)$. 
There exists an elliptic curve that passes 
through generic eight points. 
We parametrize these eight points and the curve as follows:
\begin{gather}
\begin{split}
\lefteqn{(f_1g_0+f_0g_1+\wp(2t)f_0g_0)(4\wp(2t)f_1g_1-g_3f_0g_0)}
\hspace{4cm}\\
&=\left(f_1g_1+\wp(2t)(f_1g_0+f_0g_1)+\frac{g_2}{4}f_0g_0\right)^2,
\label{eq:ellipticcurve}
\end{split}\\
p_i \colon 
\left({f_0}_i:{f_1}_i,\, {g_0}_i:{g_1}_i \right)
=\left(1:\wp(b_i+t),\,1:\wp(t-b_i) \right)
\quad(i=1,\dots,8).
\end{gather}
Here $({f_0}_i,{f_1}_i,{g_0}_i,{g_1}_i)$ depend on a variable $t$: 
\begin{equation*}
({f_0}_i,{f_1}_i,{g_0}_i,{g_1}_i)
=({f_0}_i(t),{f_1}_i(t),{g_0}_i(t),{g_1}_i(t))\quad(i=1,\dots,8).
\end{equation*}
We then define the following points:
\begin{align*}
(\bar{f_0}_i,\bar{f_1}_i,\bar{g_0}_i,\bar{g_1}_i)
&=({f_0}_i(\bar{t}),{f_1}_i(\bar{t}),{g_0}_i(\bar{t}),{g_1}_i(\bar{t}))
\quad(i=1,\dots,8),\\
(\underline{{f_0}_i},\underline{{f_1}_i},
\underline{{g_0}_i},\underline{{g_1}_i})
&=({f_0}_i(\underline{t}),{f_1}_i(\underline{t})
,{g_0}_i(\underline{t}),{g_1}_i(\underline{t}))
\quad(i=1,\dots,8),
\end{align*}
where $\bar{t}=t+\lambda,\,\underline{t}=t-\lambda,\,
\lambda=\frac{1}{2}\sum_{i=1}^8 b_i$.

The $A_0^{(1)}$-surface discrete Painlev\'e equation ($dP(A_0^{(1)})$) 
has the following trivial solution moving 
on the elliptic curve (\ref{eq:ellipticcurve}) \cite{MSY}:
\begin{equation}
({f_0}_{\mathrm{c}},{f_1}_{\mathrm{c}},{g_0}_{\mathrm{c}},{g_1}_{\mathrm{c}})
=\left(1,\,\wp(q+2t^2/\lambda+t),\,1,\,\wp(t-q-2t^2/\lambda)\right),
\label{eq:a0tri}
\end{equation}
where $q$ is a constant determined by the initial condition.
We define the following points:
\begin{align*}
(\bar{f_0}_{\mathrm{c}},\bar{f_1}_{\mathrm{c}},
\bar{g_0}_{\mathrm{c}},\bar{g_1}_{\mathrm{c}})
&=({f_0}_{\mathrm{c}}(\bar{t}),{f_1}_{\mathrm{c}}(\bar{t}),
{g_0}_{\mathrm{c}}(\bar{t}),{g_1}_{\mathrm{c}}(\bar{t}))
,\\
(\underline{{f_0}_{\mathrm{c}}},\underline{{f_1}_{\mathrm{c}}},
\underline{{g_0}_{\mathrm{c}}},\underline{{g_1}_{\mathrm{c}}})
&=({f_0}_{\mathrm{c}}(\underline{t}),{f_1}_{\mathrm{c}}(\underline{t})
,{g_0}_{\mathrm{c}}(\underline{t}),{g_1}_{\mathrm{c}}(\underline{t})).
\end{align*}

We now introduce the vectors
\begin{gather*}
\begin{split}
\lefteqn{w_{4,1}(f_0,f_1,g_0,g_1)}\hspace{1.5cm}\\
&={}^t
\left(
\begin{smallmatrix}
{f_1}^4g_1 & f_0{f_1}^3g_1 & {f_0}^2{f_1}^2g_1 & 
{f_0}^3f_1g_1 & {f_0}^4g_1 & 
{f_1}^4g_0 & f_0{f_1}^3g_0 & {f_0}^2{f_1}^2g_0 & 
{f_0}^3f_1g_0 & {f_0}^4g_0
\end{smallmatrix}
\right),
\end{split}\\
\begin{aligned}
v&=w_{4,1}(f_0,f_1,g_0,g_1), &\quad & &
\check{v}&=w_{4,1}(f_0,f_1,\bar{g_0},\bar{g_1}),\\
u&=w_{4,1}(g_0,g_1,f_0,f_1), &\quad & &
\hat{u}&=w_{4,1}(g_0,g_1,\underline{f_0},\underline{f_1}),\\
v_i&=w_{4,1}({f_0}_i,{f_1}_i,{g_0}_i,{g_1}_i), &\quad & &
\check{v}_i&=w_{4,1}({f_0}_i,{f_1}_i,\bar{g_0}_i,\bar{g_1}_i),\\ 
u_i&=w_{4,1}({g_0}_i,{g_1}_i,{f_0}_i,{f_1}_i), &\quad & &
\hat{u}_i
&=w_{4,1}({g_0}_i,{g_1}_i,\underline{{f_0}_i},\underline{{f_1}_i})
&(i=1,\dots,8),\\
v_{\mathrm{c}}&=w_{4,1}({f_0}_{\mathrm{c}},{f_1}_{\mathrm{c}},
{g_0}_{\mathrm{c}},{g_1}_{\mathrm{c}}), &\quad & &
\check{v}_{\mathrm{c}}&=w_{4,1}({f_0}_{\mathrm{c}},{f_1}_{\mathrm{c}},
\bar{g_0}_{\mathrm{c}},\bar{g_1}_{\mathrm{c}}),\\ 
u_{\mathrm{c}}&=w_{4,1}({g_0}_{\mathrm{c}},{g_1}_{\mathrm{c}},
{f_0}_{\mathrm{c}},{f_1}_{\mathrm{c}}), &\quad & &
\hat{u}_{\mathrm{c}}
&=w_{4,1}({g_0}_{\mathrm{c}},{g_1}_{\mathrm{c}},
\underline{{f_0}_{\mathrm{c}}},\underline{{f_1}_{\mathrm{c}}}).
\end{aligned}
\end{gather*}
\begin{gather*}
\begin{split}
\lefteqn{w_{3,1}(f_0,f_1,g_0,g_1)}\hspace{1cm}\\
&={}^t
\begin{pmatrix}
{f_1}^3g_1 & {f_0}{f_1}^2g_1 & 
{f_0}^2f_1g_1 & {f_0}^3g_1 & 
{f_1}^3g_0 & {f_0}{f_1}^2g_0 & 
{f_0}^2f_1g_0 & {f_0}^3g_0
\end{pmatrix},
\end{split}\\
\begin{aligned}
\phi_i&=w_{3,1}({f_0}_i,{f_1}_i,{g_0}_i,{g_1}_i), &\quad& &
\check{\phi}_i&=w_{3,1}({f_0}_i,{f_1}_i,\bar{g_0}_i,\bar{g_1}_i),\\ 
\psi_i&=w_{3,1}({g_0}_i,{g_1}_i,{f_0}_i,{f_1}_i), &\quad& &
\hat{\psi}_i
&=w_{3,1}({g_0}_i,{g_1}_i,\underline{{f_0}_i},\underline{{f_1}_i})
& \quad(i=1,\dots,8).
\end{aligned}
\end{gather*}

We describe the following theorem by using these vectors.
\begin{thm}
$dP(A_0^{(1)})$ can be written as
\begin{subequations}\label{subeqs:dp}
\begin{align}
\begin{split}
\lefteqn{
\det 
\left(
v, v_1, v_2, v_3, v_4, v_5, v_6, v_7, v_8, v_{\mathrm{c}}
\right)
\det 
\left(
\check{v}, \check{v_1}, \check{v_2}, \check{v_3}, \check{v_4}, 
\check{v_5}, \check{v_6}, \check{v_7}, \check{v_8}, \check{v_{\mathrm{c}}}
\right)
}\hspace{0.5mm}\\
&=
\det
\left(
\phi_1, \phi_2, \phi_3, \phi_4, \phi_5, \phi_6, \phi_7, \phi_8
\right)
\det
\left(
\check{\phi_1}, \check{\phi_2}, \check{\phi_3}, \check{\phi_4}, 
\check{\phi_5}, \check{\phi_6}, \check{\phi_7}, \check{\phi_8}
\right)\\
&\ {}\times 
\prod_{i=1}^{8}
\left({f_0}_{\mathrm{c}}{f_1}_i-{f_1}_{\mathrm{c}}{f_0}_i\right)\times
\left({g_0}_{\mathrm{c}}{g_1}-{g_1}_{\mathrm{c}}{g_0}\right)
\left(\bar{g_0}_{\mathrm{c}}\bar{g_1}-\bar{g_1}_{\mathrm{c}}\bar{g_0}\right)
\prod_{i=1}^8\left({f_0}_if_1-{f_1}_if_0\right),
\end{split}\label{subeqs:dp-1}\\
\begin{split}
\lefteqn{
\det 
\left(
u, u_1, u_2, u_3, u_4, u_5, u_6, u_7, u_8, u_{\mathrm{c}}
\right)
\det 
\left(
\hat{u}, \hat{u_1}, \hat{u_2}, \hat{u_3}, \hat{u_4}, 
\hat{u_5}, \hat{u_6}, \hat{u_7}, \hat{u_8}, \hat{u_{\mathrm{c}}}
\right)
}\hspace{0.5mm}\\
&=
\det
\left(
\psi_1, \psi_2, \psi_3, \psi_4, \psi_5, \psi_6, \psi_7, \psi_8
\right)
\det
\left(
\hat{\psi_1}, \hat{\psi_2}, \hat{\psi_3}, \hat{\psi_4}, 
\hat{\psi_5}, \hat{\psi_6}, \hat{\psi_7}, \hat{\psi_8}
\right)\\
&\ {}\times \prod_{i=1}^{8}
\left({g_0}_{\mathrm{c}}{g_1}_i-{g_1}_{\mathrm{c}}{g_0}_i\right)\times
\left({f_0}_{\mathrm{c}}{f_1}-{f_1}_{\mathrm{c}}{f_0}\right)
\left(\underline{{f_0}_{\mathrm{c}}}\,\underline{f_1}
-\underline{{f_1}_{\mathrm{c}}}\,\underline{f_0}\right)
\prod_{i=1}^8\left({g_0}_ig_1-{g_1}_ig_0\right).
\end{split}\label{subeqs:dp-2}
\end{align}
\end{subequations}
\end{thm}

For later use, we calculate the determinants 
$C$ and $D$ given by:
\begin{subequations}\label{subeqs:cd}
\begin{align}
\begin{split}
C&=\det
\left(
\phi_1, \phi_2, \phi_3, \phi_4, \phi_5, \phi_6, \phi_7, \phi_8
\right)
\det
\left(
\check{\phi_1}, \check{\phi_2}, \check{\phi_3}, \check{\phi_4}, 
\check{\phi_5}, \check{\phi_6}, \check{\phi_7}, \check{\phi_8}
\right)\\
&\quad{} \times
\prod_{i=1}^{8}
\left({f_0}_{\mathrm{c}}{f_1}_i-{f_1}_{\mathrm{c}}{f_0}_i\right),
\end{split}\\
\begin{split}
D&=\det
\left(
\psi_1, \psi_2, \psi_3, \psi_4, \psi_5, \psi_6, \psi_7, \psi_8
\right)
\det
\left(
\hat{\psi_1}, \hat{\psi_2}, \hat{\psi_3}, \hat{\psi_4}, 
\hat{\psi_5}, \hat{\psi_6}, \hat{\psi_7}, \hat{\psi_8}
\right)\\
&\quad{} \times
\prod_{i=1}^{8}
\left({g_0}_{\mathrm{c}}{g_1}_i-{g_1}_{\mathrm{c}}{g_0}_i\right),
\end{split}
\end{align}
\end{subequations}
for which we find:
\begin{subequations}
\begin{align}
\begin{split}
C
&=\frac{\sigma(4t)^4\sigma(4t+2\lambda)^4
\prod_{
\begin{subarray}{c}
i,j=1,\dots,8\\
i<j
\end{subarray}
}
\sigma(b_i-b_j)^2
}{\sigma(q+2t^2/\lambda+t)^{16}}\\
&\qquad{}\times
\prod_{i=1}^8
\frac{\sigma(q+2t^2/\lambda-b_i)\sigma(q+2t^2/\lambda+b_i+2t)}
{\sigma(b_i+t)^{14}\sigma(t-b_i)^2\sigma(t+\lambda-b_i)^2},
\end{split}\\
\begin{split}
D&=\frac{\sigma(4t)^4\sigma(4t-2\lambda)^4
\prod_{
\begin{subarray}{c}
i,j=1,\dots,8\\
i<j
\end{subarray}
}
\sigma(b_i-b_j)^2
}{\sigma(t-q-2t^2/\lambda)^{16}}\\
&\qquad{}\times
\prod_{i=1}^8
\frac{\sigma(q+2t^2/\lambda-b_i)\sigma(2t-q-2t^2/\lambda-b_i)}
{\sigma(t-b_i)^{14}\sigma(b_i+t)^2\sigma(b_i+t-\lambda)^2}.
\end{split}
\end{align}
\end{subequations}

\begin{rem}
We can parametrize an isomorphism class of surfaces by using the 
period mapping. 
The period mapping maps the elements of the second homology to 
$\mathbb{C}$. 

Let $\omega$ be a meromorphic 2-form on $X$ with 
$\mathrm{div}(\omega)=-D$. 
Then $\omega$ determines a period mapping 
$\hat{\chi} \colon H_2(X-D,\mathbb{Z})\to\mathbb{C}$ which sends 
$\Gamma \in H_2(X-D,\mathbb{Z})$ to $\int_{\Gamma}\omega$. 

Now, there exists a short exact sequence: 
\begin{displaymath}
0 \to H_1(D,\mathbb{Z}) \to H_2(X-D,\mathbb{Z}) \to Q(E_8^{(1)}) 
\to 0,
\end{displaymath}
where $\displaystyle{Q(E_8^{(1)})=\sum_{i=0}^8 \mathbb{Z} \alpha_i}$ 
is the root lattice of type  $E_8^{(1)}$.
So we obtain the mapping
\begin{displaymath}
\chi \colon Q(E_8^{(1)}) \to \mathbb{C} \quad \mod \ 
\hat{\chi}(H_1(D,\mathbb{Z}))
\end{displaymath}
through the period mapping $\hat{\chi}$. 
In this case, the parametrization is
\begin{equation}
\begin{gathered}
\chi(\alpha_1)=-4t,\quad \chi(\alpha_2)=b_1+b_2+2t,\quad
\chi(\alpha_i)=b_i-b_{i-1}\ (i=3,\dots,7),\\
\chi(\alpha_8)=b_2-b_1, \quad \chi(\alpha_0)=b_8-b_7. 
\end{gathered}
\end{equation}

Here $Q(E_8^{(1)})$ is realized in 
$\mathrm{Pic}(X)=H_2(X,\mathbb{Z})$. 
And $\alpha_i$'s are represented by elements of the Picard group 
as follows: 
\begin{equation}
\begin{gathered}
\alpha_1=H_1-H_0,\quad \alpha_2=H_0-E_1-E_2,\quad
\alpha_i=E_{i-1}-E_{i}\ (i=3,\dots,7),\\
\alpha_8=E_1-E_2, \quad \alpha_0=E_7-E_8.
\end{gathered}
\end{equation}
We denote the total transform of $f_1=\text{constant}\times f_0$, 
(or $g_1=\text{constant}\times g_0$) on $X$ by $H_0$ (or $H_1$ respectively) 
and the total transform of the point $p_i$ by $E_i$. 
The Picard group $\mathrm{Pic}(X)$ and canonical divisor 
$\mathcal{K}_X$ are 
\begin{displaymath}
\mathrm{Pic}(X)=\mathbb{Z} H_0+\mathbb{Z} H_1
+\sum_{i=1}^8 \mathbb{Z} E_i,\quad
\mathcal{K}_X=-2H_0-2H_1+\sum_{i=1}^8 E_i,
\end{displaymath}
where the intersection numbers of pairs of base elements are
\begin{displaymath}
H_i\cdot H_j= 1-\delta_{i,j},\quad E_i\cdot E_j= -\delta_{i,j},\quad 
H_i\cdot E_j= 0,\quad \text{where}\ 
\delta_{i,j}=
\begin{cases}
1&i=j,\\
0&i\ne j.
\end{cases}
\end{displaymath}
\end{rem}

The generators of the affine Weyl group 
$W(E_8^{(1)})=\left<w_i\ (i=0,1,\dots,8)\right>$ will act on the 
total transforms.
We give a representation of these actions that will enable us to construct 
$dP(A_0^{(1)})$. 
\begin{align*}
w_1 \colon &(H_0,H_1) \mapsto (H_1,H_0),\\
w_2 \colon &(H_1,E_1,E_2) \mapsto (H_1+H_0-E_1-E_2,H_0-E_2,H_0-E_1),\\
w_i \colon &(E_{i-1},E_i) \mapsto (E_i,E_{i-1})\quad (i=3,\dots,7),\\
w_8 \colon &(E_1,E_2) \mapsto (E_2,E_1),\\
w_0 \colon &(E_7,E_8)\mapsto(E_8,E_7).
\end{align*}

By taking a translation contained in $W(E_8^{(1)})$, 
we obtain a nonlinear difference equation. 
The translation can be described by a product of simple reflections 
$w_i$: 
\begin{gather}
dP(A_0^{(1)})=w_{1}\circ r^{-1}\circ w_{1}\circ r \colon\\
(b_i,t,f_0:f_1,g_0:g_1) \mapsto 
(b_i,t+\lambda,\bar{f_0}:\bar{f_1},\bar{g_0}:\bar{g_1})
\quad (i=1,\dots,8),\notag\\
\lambda=\frac{1}{2}\sum_{i=1}^8 b_i,\notag
\intertext{where}
\begin{split}
r&=
w_{2}\circ w_{3}\circ w_{4}\circ w_{5}\circ w_{6}\circ 
w_{7}\circ w_{0}\circ w_{8}\circ w_{3}\circ w_{4}\circ 
w_{5}\circ w_{6}\circ w_{7}\circ w_{2}\circ  \\
&\ {}\circ  
w_{3}\circ w_{4}\circ w_{5}\circ w_{6}\circ w_{8}\circ 
w_{3}\circ w_{4}\circ w_{5}\circ w_{2}\circ w_{3}\circ 
w_{4}\circ w_{8}\circ w_{3}\circ w_{2}.
\end{split}
\end{gather}

An element $r$ of the affine Weyl group acts on these 
total transforms as
\begin{equation}
r \colon (H_0,H_1,E_i) \mapsto 
(H_0,H_1+4H_0-\sum_{i=1}^8 E_i,H_0-E_{9-i}).
\end{equation}
Equation (\ref{subeqs:dp-1}) describes exactly this action. 
The following gives a presentation of the action of $r$ on coordinates and 
parameters:
\begin{equation}
\begin{split}
\lefteqn{r \colon (f_0:f_1,\,g_0:g_1,{f_0}_i:{f_1}_i,\,{g_0}_i:{g_1}_i)}
\hspace{3cm}\\
&\mapsto 
(\check{f_0}:\check{f_1},\,\check{g_0}:\check{g_1},
\check{f_0}_{i}:\check{f_1}_{i},\,\check{g_0}_{i}:\check{g_1}_{i})\\
&=(f_0:f_1,\,\bar{g_0}:\bar{g_1},
{f_0}_{(9-i)}:{f_1}_{(9-i)},\,\bar{g_0}_{(9-i)}:\bar{g_1}_{(9-i)}).
\end{split}
\end{equation}
For example, if we impose 
$\check{g_0}_{\mathrm{c}}\check{g_1}-\check{g_1}_{\mathrm{c}}\check{g_0}=0$
in the equation (\ref{subeqs:dp-1}), i.e. 
we impose $\bar{g_0}_{\mathrm{c}}\bar{g_1}-\bar{g_1}_{\mathrm{c}}\bar{g_0}=0$ 
on $(f_0:f_1,\,\bar{g_0}:\bar{g_1})$ in the equation,
we obtain 
$\det 
\left(
v, v_1, v_2, v_3, v_4, v_5, v_6, v_7, v_8, v_{\mathrm{c}}
\right)=0$ on $(f_0:f_1,\,g_0:g_1)$ as an equivalent condition.
This expresses the action $r\colon H_1 \mapsto H_1+4H_0-\sum_{i=1}^8 E_i$.
If we impose $\check{f_0}_i\check{f_1}-\check{f_1}_i\check{f_0}=0,\,
\check{g_0}_i\check{g_1}-\check{g_1}_i\check{g_0}=0$ 
in the equation (\ref{subeqs:dp-1}), i.e.\
we impose 
${f_0}_{(9-i)}{f_1}-{f_1}_{(9-i)}{f_0}=0,\,
\bar{g_0}_{(9-i)}\bar{g_1}-\bar{g_1}_{(9-i)}\bar{g_0}=0$ 
on $(f_0:f_1,\,\bar{g_0}:\bar{g_1})$ in the equation,
we obtain ${f_0}_{(9-i)}f_1-{f_1}_{(9-i)}f_0=0$ 
on $(f_0:f_1,\,g_0:g_1)$.
This expresses $r\colon E_i \mapsto H_0-E_{9-i}$.

If we decide the form of the equation in this way, 
an undetermined constant will remain.
However the condition that the trivial solution (\ref{eq:a0tri}) 
(for arbitrary $q$) satisfies the equation 
completely determines the form of (\ref{subeqs:dp}).

\begin{rem}
We presented the $dP(A_0^{(1)})$ in \cite{MSY} as follows. 
That is, in this section we have rewritten this expression.
This system is equivalent to $\mathrm{ell.} P$ derived in \cite{S}. 
The following $2\times 2$ matrices represent PGL(2)-action 
on $\mathbb{P}^1$, i.e.,
$w=
\left(
\begin{smallmatrix}
a & b\\
c & d
\end{smallmatrix}
\right)
z$
means 
$w=(az+b)/(cz+d)$.
The $A_0^{(1)}$-surface discrete Painlev\'e equation is 
the following difference system for unknown functions $f(t),\,g(t)$:
\begin{subequations}
\begin{align}
\begin{split}
\bar{g}&=
M\left(f,c_7,c_8,t-\frac{1}{4}\sum_{i=1}^6 c_i\right)
M\left(f,c_5,c_6,t-\frac{1}{4}\sum_{i=1}^4 c_i\right)\\
&\quad {}\times
M\left(f,c_3,c_4,t-\frac{1}{4}(c_1+c_2)\right)
M(f,c_1,c_2,t)\, g,
\end{split} \label{eq:ellipticp1}\\
\begin{split}
\underline{f}&=
M\left(g,d_7,d_8,t-\frac{1}{4}\sum_{i=1}^6 d_i\right)
M\left(g,d_5,d_6,t-\frac{1}{4}\sum_{i=1}^4 d_i\right)\\
&\quad {}\times 
M\left(g,d_3,d_4,t-\frac{1}{4}(d_1+d_2)\right)
M(g,d_1,d_2,t)\, f, 
\end{split}\label{eq:ellipticp2}
\end{align}\label{subeqs:epold}
\end{subequations}
where $\bar{g}=g(t+\lambda)$,
$\underline{f}=f(t-\lambda)$
and 
\begin{equation}
\begin{split}
\lefteqn{M(h,\kappa_1,\kappa_2,s)}\hspace{1mm}\\
&=
\begin{pmatrix}
-\wp(2s-\frac{-\kappa_1+\kappa_2}{2}) & 
\wp(2s-\frac{\kappa_1-\kappa_2}{2})\\
-1 & 1
\end{pmatrix}\\
& {}\times
\begin{pmatrix}
(h-\wp(\kappa_2))(\wp(2s)-\wp(2s-\kappa_2))
(\wp(2s-\frac{\kappa_1+\kappa_2}{2})
-\wp(2s-\frac{\kappa_1-\kappa_2}{2})) \quad 0\\
0 \quad (h-\wp(\kappa_1))(\wp(2s)-\wp(2s-\kappa_1))
(\wp(2s-\frac{\kappa_1+\kappa_2}{2})
-\wp(2s-\frac{-\kappa_1+\kappa_2}{2}))
\end{pmatrix}\\
& {}\times
\begin{pmatrix}
1 & -\wp(2s-\kappa_1)\\
1 & -\wp(2s-\kappa_2)
\end{pmatrix}.
\end{split} \label{ellipticM}
\end{equation}
Here $b_i\ (i=1,\dots,8)$ are constant parameters and we set
$\lambda=\frac{1}{2}\sum_{i=1}^{8}b_i,c_i=b_i+t,d_i=t-b_i$.
Note that we will regard $(f(t),g(t))$ as inhomogeneous coordinates 
of $\mathbb{P}^1\times\mathbb{P}^1$.
\end{rem}

\section{$A_0^{(1)*}$-surface}\label{a0*}
We construct the $A_0^{(1)*}$-surface by blowing up 
$\mathbb{P}^1\times\mathbb{P}^1$ at eight points. 
These eight points and a curve through these points are 
\begin{gather}
{f_1}^2{g_0}^2+{f_0}^2{g_1}^2
-\left(t^2+\frac{1}{t^2}\right)f_0f_1g_0g_1
+\left(t^2-\frac{1}{t^2}\right)^2{f_0}^2{g_0}^2=0,\label{eq:a0*curve}\\
p_i \colon \left({f_0}_i:{f_1}_i,\, {g_0}_i:{g_1}_i \right)
=\left(1:b_it+\frac{1}{b_it},
1:\frac{t}{b_i}+\frac{b_i}{t} \right)
\quad(i=1,\dots,8).
\end{gather}

The $A_0^{(1)*}$-surface discrete Painlev\'e equation ($dP(A_0^{(1)*})$) 
has the following trivial solution moving 
on the curve (\ref{eq:a0*curve}) \cite{MSY}:
\begin{multline}
({f_0}_{\mathrm{c}},{f_1}_{\mathrm{c}},{g_0}_{\mathrm{c}},{g_1}_{\mathrm{c}})\\
=
\left(1,
tq\exp\left(\tfrac{2(\log t)^2}{\log \lambda}\right)
+\tfrac{1}{tq\exp\left(\frac{2(\log t)^2}{\log \lambda}\right)},
1,\tfrac{t}{q\exp\left(\frac{2(\log t)^2}{\log \lambda}\right)}
+\tfrac{q\exp\left(\frac{2(\log t)^2}{\log \lambda}\right)}{t}\right),
\label{eq:a0*tri}
\end{multline}
where $q$ is a constant determined by the initial condition.

Using these expressions we formulate the theorem:
\begin{thm}
$dP(A_0^{(1)*})$ can be written in the form (\ref{subeqs:dp}).
\end{thm}

Let $q=0$, then the trivial solution (\ref{eq:a0*tri}) is
\begin{equation}
({f_0}_{\mathrm{c}},{f_1}_{\mathrm{c}},{g_0}_{\mathrm{c}},{g_1}_{\mathrm{c}})
=\left(0,1,0,1 \right),
\end{equation}
and $C,D$ (\ref{subeqs:cd}) are
\begin{subequations}
\begin{align}
C&=\left(t^2-\frac{1}{t^2}\right)^4
\left(t\bar{t}-\frac{1}{t\bar{t}}\right)^4
\prod_{
\begin{subarray}{c}
i,j=1,\dots,8\\
i<j
\end{subarray}
}
\left(b_i-b_j\right)^2 \Bigg/ \prod_{i=1}^8 {b_i}^7,\\
D&=\left(t^2-\frac{1}{t^2}\right)^4
\left(t\underline{t}-\frac{1}{t\underline{t}}\right)^4
\prod_{
\begin{subarray}{c}
i,j=1,\dots,8\\
i<j
\end{subarray}
}
\left(b_i-b_j\right)^2 \Bigg/ \prod_{i=1}^8 {b_i}^7.
\end{align}
\end{subequations}

\section{$A_0^{(1)**}$-surface}
We construct the $A_0^{(1)**}$-surface by blowing up 
$\mathbb{P}^1\times\mathbb{P}^1$ at eight points. 
These eight points and a curve on which these points lie are 
as follows:
\begin{gather}
(f_1g_0-f_0g_1)^2-8t^2(f_0f_1{g_0}^2+{f_0}^2g_0g_1)
+16t^4{f_0}^2{g_0}^2=0,\label{eq:a0**curve}\\
p_i \colon \left({f_0}_i:{f_1}_i,\, {g_0}_i:{g_1}_i \right)
=\left(1:(b_i+t)^2,1:(t-b_i)^2 \right)
\quad(i=1,\dots,8),
\end{gather}

The $A_0^{(1)**}$-surface discrete Painlev\'e equation ($dP(A_0^{(1)**})$) 
has the following trivial solution moving 
on the curve (\ref{eq:a0**curve}) \cite{MSY}:
\begin{equation}
({f_0}_{\mathrm{c}},{f_1}_{\mathrm{c}},{g_0}_{\mathrm{c}},{g_1}_{\mathrm{c}})
=\left(1,\,\left(q+2t^2/\lambda+t\right)^2,\,
1,\,\left(t-q-2t^2/\lambda\right)^2\right),\label{eq:a0**tri}
\end{equation}
where $q$ is a constant determined by initial condition.

\begin{thm}
$dP(A_0^{(1)**})$ can be written in the form (\ref{subeqs:dp}).
\end{thm}

If $q=\infty$ the trivial solution (\ref{eq:a0**tri}) is
\begin{equation}
({f_0}_{\mathrm{c}},{f_1}_{\mathrm{c}},
{g_0}_{\mathrm{c}},{g_1}_{\mathrm{c}})
=\left(0,1,0,1 \right)
\end{equation}
and $C,D$ (\ref{subeqs:cd}) are
\begin{subequations}
\begin{align}
C&=4096\,t^{4}\left(t+\bar{t}\right)^4
\prod_{
\begin{subarray}{c}
i,j=1,\dots,8\\
i<j
\end{subarray}
}
\left(b_i-b_j\right)^2,\\
D&=4096\,t^{4}\left(t+\underline{t}\right)^4
\prod_{
\begin{subarray}{c}
i,j=1,\dots,8\\
i<j
\end{subarray}
}
\left(b_i-b_j\right)^2.
\end{align}
\end{subequations}

\section{$A_1^{(1)}$-surface}
We construct the $A_1^{(1)}$-surface by blowing up 
$\mathbb{P}^1\times\mathbb{P}^1$ at eight points. 
These eight points and a curve through these points are 
\begin{gather}
(f_1g_1-t^2f_0g_0)(f_1g_1-f_0g_0)=0,\\
p_i \colon \left({f_0}_i:{f_1}_i,\, {g_0}_i:{g_1}_i \right)
=
\begin{cases}
\left(1:b_it,\,1:\frac{t}{b_i} \right)
\quad&(i=1,\dots,4),\\
\left(1:b_i,\,1:\frac{1}{b_i} \right)
\quad&(i=5,\dots,8).
\end{cases}
\end{gather}

The $A_1^{(1)}$-surface discrete Painlev\'e equation ($dP(A_1^{(1)})$) 
has the following trivial solution.
\begin{equation*}
({f_0}_{\mathrm{c}},{f_1}_{\mathrm{c}},{g_0}_{\mathrm{c}},{g_1}_{\mathrm{c}})
=\left(0,\,1,\,1,\,0\right).
\end{equation*}

\begin{thm}
$dP(A_1^{(1)})$ can be written in the form (\ref{subeqs:dp}).
\end{thm}

$C,D$ (\ref{subeqs:cd}) are
\begin{subequations}
\begin{align}
C&=t^8\left(1-t^2\right)^4\left(1-t\bar{t}\right)^4
\prod_{
\begin{subarray}{c}
i,j=1,\dots,4\\
i<j
\end{subarray}
}
\left(b_i-b_j\right)^2
\prod_{
\begin{subarray}{c}
i,j=5,\dots,8\\
i<j
\end{subarray}
}
\left(b_i-b_j\right)^2
\Bigg/
\prod_{i=1}^8
b_i,\\
D&=t^{12}\left(1-t^2\right)^4
\left(1-t\underline{t}\right)^4
\prod_{
\begin{subarray}{c}
i,j=1,\dots,4\\
i<j
\end{subarray}
}
\left(b_i-b_j\right)^2
\prod_{
\begin{subarray}{c}
i,j=5,\dots,8\\
i<j
\end{subarray}
}
\left(b_i-b_j\right)^2
\Bigg/
\prod_{i=1}^8
{b_i}^6.
\end{align}
\end{subequations}

\section{$A_1^{(1)*}$-surface}
We construct the $A_1^{(1)*}$-surface by blowing up 
$\mathbb{P}^1\times\mathbb{P}^1$ at eight points. 
These eight points and a curve passing through these points are 
\begin{gather}
(f_1g_0+f_0g_1-2tf_0g_0)(f_1g_0+f_0g_1)=0,\\
p_i \colon \left({f_0}_i:{f_1}_i,\, {g_0}_i:{g_1}_i \right)
=
\begin{cases}
\left(1:b_i+t,\,1:t-b_i \right)
\quad&(i=1,\dots,4),\\
\left(1:b_i,\,1:-b_i \right)
\quad&(i=5,\dots,8).
\end{cases}
\end{gather}

The $A_1^{(1)*}$-surface discrete Painlev\'e equation ($dP(A_1^{(1)*})$) 
has the following trivial solution.
\begin{equation*}
({f_0}_{\mathrm{c}},{f_1}_{\mathrm{c}},{g_0}_{\mathrm{c}},{g_1}_{\mathrm{c}})
=\left(0,\,1,\,0,\,1\right)
\end{equation*}
and we have the following theorem:
\begin{thm}
$dP(A_1^{(1)*})$ can be written in the form (\ref{subeqs:dp}).
\end{thm}

$C,D$ (\ref{subeqs:cd}) are
\begin{subequations}
\begin{align}
C&=16t^4\left(t+\bar{t}\right)^4
\prod_{
\begin{subarray}{c}
i,j=1,\dots,4\\
i<j
\end{subarray}
}
\left(b_i-b_j\right)^2
\prod_{
\begin{subarray}{c}
i,j=5,\dots,8\\
i<j
\end{subarray}
}
\left(b_i-b_j\right)^2,\\
D&=16t^4\left(t+\underline{t}\right)^4
\prod_{
\begin{subarray}{c}
i,j=1,\dots,4\\
i<j
\end{subarray}
}
\left(b_i-b_j\right)^2
\prod_{
\begin{subarray}{c}
i,j=5,\dots,8\\
i<j
\end{subarray}
}
\left(b_i-b_j\right)^2.
\end{align}
\end{subequations}

\section{$A_2^{(1)}$-surface}
We construct $A_2^{(1)}$-surface by blowing up 
$\mathbb{P}^1\times\mathbb{P}^1$ at eight points. 
These eight points and a curve on which these points lie are 
\begin{gather}
f_0g_0(f_1g_1-f_0g_0)=0,\\
p_i \colon 
\left({f_0}_i:{f_1}_i,\, {g_0}_i:{g_1}_i \right)
=
\begin{cases}
\left(1:b_it,\,0:1 \right)
\quad&(i=1,2),\\
\left(0:1,\,1:\frac{t}{b_i} \right)
\quad&(i=3,4),\\
\left(1:b_i,\,1:\frac{1}{b_i} \right)
\quad&(i=5,\dots,8).
\end{cases}
\end{gather}

The $A_2^{(1)}$-surface discrete Painlev\'e equation ($dP(A_2^{(1)})$) 
has the trivial solution
\begin{equation*}
({f_0}_{\mathrm{c}},{f_1}_{\mathrm{c}},{g_0}_{\mathrm{c}},{g_1}_{\mathrm{c}})
=\left(0,\,1,\,1,\,0\right).
\end{equation*}

If we write $dP(A_2^{(1)})$ in the form (\ref{subeqs:dp}),
then both sides of (\ref{subeqs:dp-1}) are $0$. 
However we can derive the following expression from the form 
of $dP(A_1^{(1)})$.

\begin{thm}
$dP(A_2^{(1)})$ can be written as
\begin{subequations}
\begin{align}
\begin{split}
\lefteqn{
\det 
\left(
v, v_1, v_2, v_3, v_4, v_5, v_6, v_7, v_8, \mathrm{v_{\mathrm{c}}}
\right)
\det 
\left(
\check{v}, \check{v_1}, \check{v_2}, \check{v_3}, \check{v_4}, 
\check{v_5}, \check{v_6}, \check{v_7}, \check{v_8}, 
\check{\mathrm{v_{\mathrm{c}}}}
\right)
}\hspace{2cm}\\
&=
t^3\bar{t}^3(b_1-b_2)^2(b_3-b_4)^2
\prod_{
\begin{subarray}{c}
i,j=5,\dots,8\\
i<j
\end{subarray}}
\left(b_i-b_j\right)^2
\Big/{b_3}^3{b_4}^3\\
&\quad{}\times \left({g_0}_{\mathrm{c}}{g_1}-{g_1}_{\mathrm{c}}{g_0}\right)
\left(\bar{g_0}_{\mathrm{c}}\bar{g_1}-\bar{g_1}_{\mathrm{c}}\bar{g_0}\right)
\prod_{i=1}^8\left({f_0}_if_1-{f_1}_if_0\right),
\end{split}\\
\begin{split}
\lefteqn{
\det 
\left(
u, u_1, u_2, u_3, u_4, u_5, u_6, u_7, u_8, u_{\mathrm{c}}
\right)
\det 
\left(
\hat{u}, \hat{u_1}, \hat{u_2}, \hat{u_3}, \hat{u_4}, 
\hat{u_5}, \hat{u_6}, \hat{u_7}, \hat{u_8}, \hat{u_{\mathrm{c}}}
\right)
}\hspace{2cm}\\
&=t^9\underline{t}
(b_1-b_2)^2(b_3-b_4)^2
\prod_{
\begin{subarray}{c}
i,j=5,\dots,8\\
i<j
\end{subarray}}
\left(b_i-b_j\right)^2
\Bigg/
\prod_{i=3}^8 {b_i}^5\\
&\quad{}\times \left({f_0}_{\mathrm{c}}{f_1}-{f_1}_{\mathrm{c}}{f_0}\right)
\left(\underline{{f_0}_{\mathrm{c}}}\,\underline{f_1}
-\underline{{f_1}_{\mathrm{c}}}\,\underline{f_0}\right)
\prod_{i=1}^8\left({g_0}_ig_1-{g_1}_ig_0\right),
\end{split}
\end{align}
\end{subequations}
where
\begin{equation*}
\mathrm{v_{\mathrm{c}}}=\check{\mathrm{v_{\mathrm{c}}}}={}^t
\begin{pmatrix}
0 & 1 & 0 & 0 & 0 & 0 & 0 & 0 & 0 & 0
\end{pmatrix}.
\end{equation*}
\end{thm}

\section{$A_3^{(1)}$-surface}\label{a3}
We construct the $A_3^{(1)}$-surface by blowing up 
$\mathbb{P}^1\times\mathbb{P}^1$ at eight points. 
These eight points and a curve through them are 
\begin{gather}
f_0f_1g_0g_1=0,\\
p_i \colon 
\left({f_0}_i:{f_1}_i,\, {g_0}_i:{g_1}_i \right)
=
\begin{cases}
\left(1:b_i,0:1 \right)\quad&(i=1,2),\\
\left(0:1,1:\frac{1}{b_i} \right)\quad&(i=3,4),\\
\left(1:b_it,1:0 \right)\quad&(i=5,6),\\
\left(1:0,1:\frac{t}{b_i} \right)\quad&(i=7,8).
\end{cases}
\end{gather}

The $A_3^{(1)}$-surface discrete Painlev\'e equation ($dP(A_3^{(1)})$) 
has the trivial solution.
\begin{equation*}
({f_0}_{\mathrm{c}},{f_1}_{\mathrm{c}},{g_0}_{\mathrm{c}},{g_1}_{\mathrm{c}})
=\left(0,\,1,\,1,\,0\right),
\end{equation*}
for which we have
\begin{thm}
$dP(A_3^{(1)})$ can be written as
\begin{subequations}\label{subeqs:a3}
\begin{align}
\begin{split}
\lefteqn{
\det 
\left(
v, v_1, v_2, v_3, v_4, v_5, v_6, v_7, v_8, \mathrm{v_{\mathrm{c}}}
\right)
\det 
\left(
\check{v}, \check{v_1}, \check{v_2}, \check{v_3}, \check{v_4}, 
\check{v_5}, \check{v_6}, \check{v_7}, \check{v_8}, 
\check{\mathrm{v_{\mathrm{c}}}}
\right)
}\hspace{1mm}\\
&=\frac{1}{(b_1-b_2)^2(b_3-b_4)^2(b_5-b_6)^2(b_7-b_8)^2 t^7\bar{t}}
\left(\frac{b_3b_4b_7b_8}{b_1b_2b_5b_6}\right)^2\\
&\quad{}\times 
g_1\bar{g}_1(f_1-b_1f_0)(f_1-b_2f_0){f_0}^2(f_1-b_5tf_0)(f_1-b_6tf_0){f_1}^2,
\end{split}\\
\begin{split}
\lefteqn{
\det 
\left(
u, u_1, u_2, u_3, u_4, u_5, u_6, u_7, u_8, \mathrm{u_{\mathrm{c}}}
\right)
\det 
\left(
\hat{u}, \hat{u_1}, \hat{u_2}, \hat{u_3}, \hat{u_4}, 
\hat{u_5}, \hat{u_6}, \hat{u_7}, \hat{u_8}, \hat{\mathrm{u_{\mathrm{c}}}}
\right)
}\hspace{1mm}\\
&=
\frac{{b_3}^4{b_4}^4{b_7}^6{b_8}^6}
{(b_1-b_2)^2(b_3-b_4)^2(b_5-b_6)^2(b_7-b_8)^2 t^{11}\bar{t}}\\
&\quad{}\times 
f_0\underline{f_0}
{g_0}^2(g_1-1/b_3g_0)(g_1-1/b_4g_0){g_1}^2(g_1-t/b_7g_0)(g_1-t/b_8g_0),
\end{split}
\end{align}
\end{subequations}
where
\begin{equation*}
\mathrm{u_{\mathrm{c}}}=\hat{\mathrm{u_{\mathrm{c}}}}=
{}^t
\begin{pmatrix}
0 & 0 & 0 & 0 & 0 & 0 & 0 & 0 & 1 & 0
\end{pmatrix}.
\end{equation*}
\end{thm}

If we expand the determinant in (\ref{subeqs:a3}),
we obtain 
\begin{subequations}
\begin{align}
\frac{g_1\bar{g}_1}{g_0\bar{g}_0}
&=\frac{(f_1-b_5tf_0)(f_1-b_6tf_0)}{(f_1-b_1f_0)(f_1-b_2f_0)},\\
\frac{f_1\underline{f_1}}{f_0\underline{f_0}}
&=\frac{(g_1-t/b_7g_0)(g_1-t/b_8g_0)}{(g_1-1/b_3g_0)(g_1-1/b_4g_0)}.
\end{align}
\end{subequations}
This is exactly $q$-$\mathrm{P}_\mathrm{VI}$ \cite{JS}.

\section{Discussion}
In this paper we presented a new representation of discrete 
Painlev\'e equations. 
Up to now, 
the complexity of some of the difference Painlev\'e 
equations prevented us from studying their properties.
It is our hope that these new forms of these equations,
especially for 
$dP(A_0^{(1)})$, $dP(A_0^{(1)*})$, and $dP(A_0^{(1)**})$
which have Weyl groups of type $E_8^{(1)}$,
will prove useful in such analysis.

\bigskip
\noindent
\textit{Acknowledgement.} 
The author would like to thank K. Okamoto and H. Sakai 
for discussions and advice. 
The author is also grateful to R. Willox for useful comments.

\bibliographystyle{plain}

\end{document}